\documentclass[12pt, letterpaper]{article}

\AtBeginDocument{
  \providecommand\BibTeX{{
    \normalfont B\kern-0.5em{\scshape i\kern-0.25em b}\kern-0.8em\TeX}}}

\usepackage{geometry}
\newgeometry{vmargin={25mm}, hmargin={25mm,25mm}}

\usepackage{physics}
\usepackage{xcolor}
\usepackage{pifont} 
\usepackage{colortbl}
\usepackage{tabularx}
\usepackage{amsmath}
\usepackage{mathtools}
\usepackage{caption}
\usepackage{subcaption}
\usepackage{booktabs} 

\usepackage{listings}
\usepackage{graphicx}
\usepackage{multicol}
\usepackage{multirow}
\usepackage{ctable}	
\usepackage{url}
\usepackage{float}
\usepackage{upgreek}
\usepackage{algorithmicx}
\usepackage{algorithm}
\usepackage{algpseudocode}
\usepackage{enumitem}
\usepackage{authblk}
\usepackage{cite}

\usepackage{hyphenat}
\definecolor{mGreen}{rgb}{0,0.6,0}
\definecolor{mGray}{rgb}{0.5,0.5,0.5}
\definecolor{mPurple}{rgb}{0.58,0,0.82}
\definecolor{mOrange}{rgb}{0.9,0.4,0.1}
\definecolor{backgroundColour}{rgb}{0.95,0.95,0.92}

\lstdefinestyle{CStyle}{
    backgroundcolor=\color{backgroundColour},   
    commentstyle=\color{mGreen},
    keywordstyle=\color{magenta},
    numberstyle=\tiny\color{mGray},
    stringstyle=\color{mPurple},
    basicstyle=\scriptsize,
    breakatwhitespace=false,         
    breaklines=true,                 
    captionpos=b,                    
    keepspaces=true,                 
    numbers=left,                    
    numbersep=5pt,                  
    showspaces=false,                
    showstringspaces=false,
    showtabs=false,                  
    tabsize=2,
    frame=single,
    rulecolor=\color{black},
    language=C
}

\newcommand{\upcid}{$^\ddagger$}
\newcommand{\bscid}{$^\dagger$}

\begin{document}

\title{\color{black}An Automotive Case Study on the Limits \\ of Approximation for Object Detection}
\author[\bscid, \upcid]{Martí Caro}
\author[\bscid]{Hamid Tabani}
\author[\bscid]{Jaume Abella}
\author[\bscid, \upcid]{Francesc Moll}
\author[\upcid]{Enric Morancho}
\author[\bscid, \upcid]{Ramon Canal}
\author[\upcid]{Josep Altet}
\author[\upcid]{Antonio Calomarde}
\author[\bscid]{Francisco J. Cazorla}
\author[\upcid]{Antonio Rubio}
\author[\bscid, \upcid]{Pau Fontova}
\author[\bscid, \upcid]{Jordi Fornt}
\affil[\bscid]{Barcelona Supercomputing Center (BSC)}
\affil[\upcid]{Universitat Polit\`{e}cnica de Catalunya (UPC)}

\date{}

\maketitle

\begin{abstract}
The accuracy of camera-based object detection (CBOD) built upon deep learning is often evaluated against the real objects in frames only. However, such simplistic evaluation ignores the fact that many unimportant objects are small, distant, or background, and hence, their misdetections have less impact than those for closer, larger, and foreground objects in domains such as autonomous driving. Moreover, sporadic misdetections are irrelevant since confidence on detections is typically averaged across consecutive frames, and detection devices (e.g. cameras, LiDARs) are often redundant, thus providing fault tolerance. 

This paper exploits such intrinsic fault tolerance of the CBOD process, and assesses {\color{black}in an automotive case study} to what extent CBOD can tolerate approximation coming from multiple sources such as lower precision arithmetic, approximate arithmetic units, and even random faults due to, for instance, low voltage operation. We show that the accuracy impact of those sources of approximation is within 1\% of the baseline even when considering the three approximate domains simultaneously, and hence, multiple sources of approximation can be exploited to build highly efficient accelerators for CBOD in cars.
\end{abstract}

\section{Introduction}\label{sec:introduction}

Systems based on Artificial Intelligence are becoming ubiquitous these days for a variety of applications across domains. Those based on deep learning are particularly popular for object detection, and hardware and software designs to achieve increasingly high accuracy and confidence on the predictions within limited time bounds (e.g. to process images from cameras at a high rate) are continuously improved.

Camera-based object detection (CBOD) building upon {\color{black}deep learning} has been deployed in a variety of applications across multiple domains{\color{black}, some of them with safety requirements, and thus with limitations in the level of errors that can be tolerated. This is, for instance, the case of CBOD in Autonomous Driving (AD), where {\color{black}accurate} object detection is mandatory to guarantee safe progress towards the destination.} For instance, \emph{You Only Look Once} (YOLO)~\cite{YOLOv2,YOLOv3} is an award-winning popular and efficient real-time CBOD system already used {\color{black}in industrial autonomous driving systems such as NVIDIA Drive~\cite{nvidia-drive} and Baidu's} Apollo~\cite{apollo}, a popular industrial-quality autonomous driving software framework used in several prototype vehicles (including autonomous trucks and robotaxies).

{\color{black}YOLO implements a computationally-intensive function based on a Convolutional Neural Network. It is developed and trained based on IEEE754 floating-point 32-bit arithmetic and has been proven highly effective and efficient.}

For each frame (image), YOLO delivers the list of objects detected with their individual \emph{\textbf{confidence levels}}. {\color{black}By default,} only objects with detection confidence above 50\% are regarded as real objects for YOLO. {\color{black}{\color{black}Analogous thresholds -- even identical -- are} considered and implemented accordingly in frameworks that employ YOLO's architecture, such as Apollo.}

CBOD in general, and YOLO in particular, have a stochastic nature due to building on neural networks{\color{black} \cite{intro-cnns}}, and by delivering results with associated confidence levels. Such stochastic nature has often been leveraged to use reduced-precision arithmetic (e.g. {\color{black}floating-point 16-bit} instead of {\color{black}floating-point 32-bit}), to reduce {\color{black}computation costs while maintaining accuracy and, in some cases, at the expense of negligible accuracy loss}. 

{\color{black}While the effectiveness of object detection can be assessed at the granularity of an image, where we can tell whether objects have been detected, identifying false positives and false negatives, the semantic implications of false positives/negatives are irrelevant at the granularity of individual images for automotive object detection {\color{black}if decisions are taken averaging results across multiple frames, as it is the case of YOLO}. Instead, detections are {\color{black}combined} across multiple {\color{black}continuous} frames captured from cameras, since only when an object is detected across multiple frames -- not necessarily strictly consecutive frames -- {\color{black}with sufficiently high confidence} is regarded as a true object. For instance, an object detected in a single frame, but not in the neighbouring ones {\color{black}can be} simply disregarded. Analogously, missing to detect {\color{black}with sufficient confidence} an object in a single frame out of a sequence of frames where it is detected {\color{black}is very unlikely to} change the outcome, and hence the object is regarded as a real object.}

AD frameworks leverage detections across multiple frames and across multiple sensors (cameras, LiDARs, and radars) {\color{black}for the sake of robustness}. Thus, a false positive/negative in a frame causes no semantic error on its own.
This brings an additional dimension for error tolerance that can be exploited to further reduce hardware complexity.

This paper {\color{black}performs a case study for} the trade-off between accuracy and complexity for CBOD in the context of AD systems. In particular, we show {\color{black}for the first time in an automotive case study} that several approximation domains can be leveraged simultaneously {\color{black}with negligible impact on accuracy}:

\begin{enumerate}[label=\Alph*), topsep=8pt]
	\item Reduced precision arithmetic (e.g. {\color{black}floating-point 16-bit}), in line with existing literature.
	\item Approximate arithmetic, such as approximate additions and multiplications, where results for some input operands may be systematically inaccurate.
	\item {\color{black}Random faults due to, for instance,} low-voltage operation, which may make some operations produce arbitrarily erroneous results sporadically.
\end{enumerate}

Hence, this work shows that opportunities exist for the future design of lower-power higher-frequency AD-specific CBOD accelerators by implementing low-voltage reduced-preci-sion approximate arithmetic, without impacting on the semantics of the whole object-detec-tion process.

{\color{black}The rest of the paper is organized as follows. Related work is provided in Section~\ref{sec:related}. Section~\ref{sec:back} provides background on object detection systems as well as state-of-the-art on different approaches to improve the efficiency of {\color{black}deep learning} models. Section~\ref{sec:analysis} presents the different approximation domains and the expected impact on the CBOD outcome. Section~\ref{sec:evaluation} presents and evaluates the case study. Finally, Section~\ref{sec:concl} summarizes this work.}

\section{Related Work}
\label{sec:related}
{\color{black}
Achieving efficient object detection has been the target of multiple works from different angles, including the reduction of the amount of data fetched and operated, and reducing the cost to fetch and operate such data, being both sets of solutions orthogonal and complementary. In the former area, we find works performing real-time video segmentation to compress spatio-temporal redundancy within and across frames~\cite{reviewer3-1}, and devising new CNN architectures for low-cost salient object detection~\cite{reviewer3-2}, and enabling CNN model compaction~\cite{reviewer3-3}. Those works ultimately lead to requiring less data to be fetched from memory, and to performing a lower number of operations on the data fetched. In the latter area, we find works related to using alternative data representations, approximate arithmetics and low voltage operation, which we discuss next. 
}

There is a plethora of related works in the area of reduced precision for neural networks. The work from Tang et al. \cite{mlpat} proposes MLPAT, a power, area, and timing model framework for machine learning accelerators. The work explores the design space of precision and architecture tradeoffs for autonomous driving accelerators assessing the accuracy, power, and area of {\color{black}floating-point 16-bit}, {\color{black}brain floating-point 16-bit}, {\color{black}integer 16-bit}, and {\color{black}integer 8-bit} on a TPU-v1 \cite{10.1145/3079856.3080246}, an ASIC developed by Google. Wu et al. \cite{wu2020low} propose a custom low-precision (8-bit) floating-point quantization method for FPGA-based acceleration without re-training and study their approach on VGG16 \cite{zhang2015accelerating} and tiny-YOLO \cite{7780460}. In our work, we use the full version of YOLO v3 \cite{YOLOv3} and the standard floating-point formats. Mellempudi et al. \cite{mellempudi2019mixed} propose the use of 8-bit floating-point representation for weights, activations, errors, and gradients, although the work focuses on training a DNN.

Several works study other possible number formats, such as fixed-point, during training and inference \cite{7011421}, \cite{Carmichael_2019}, \cite{10.5555/3045390.3045690}, \cite{10.5555/3045118.3045303}, concluding that it is feasible using alternative number format representations in the context of CNNs without significant accuracy impact. However, in practice, it is currently more feasible to use the standard floating-point number representation, since the hardware is highly optimized, while other formats may require further optimization.

Approximate arithmetic applied to CNNs has also been studied in several works. The work from Hammad et al. \cite{9313992} studies the use of a dynamically configurable approximate multiplier for CNN inference on the VGG19 \cite{simonyan2015deep}, Xception \cite{chollet2017xception}, and DenseNet201 \cite{densenet} networks using the ImageNetV2 dataset \cite{recht2019imagenet}. Ibrahim et al. \cite{8617877} discuss the use of approximate computing by means of employing approximate multipliers and adders for machine learning. Further related works can be found on the comprehensive survey from Manikandan~\cite{Manikandan2020ApproximationCT}. To the best of our knowledge, there are no related works focusing on approximate arithmetic for CBOD in AD systems, and other works that study approximate arithmetic employ a lower complexity network. In our work, we have employed the {\color{black}SFA (simplified full adder) based adder~\cite{approxADD}} and a configuration of the {\color{black}UDM (underdesigned multiplier)~\cite{approxMUL}} as illustrative examples of approximate arithmetic units. Other works study different approximate units. 

{\color{black}The impact of faults on CNNs has been analyzed from different angles.}
Some other works study the impact of aggressive low voltage operations applied to CNNs. Salami et al. \cite{9153393} perform an experimental study of the power-performance-accuracy characteristics of CNN accelerators with aggressive reduced supply voltage capability implemented in real FPGAs, although the study is not focused on the AD domain. Chandramoorthy~\cite{8675205} also performs an evaluation of low-voltage operation, {\color{black}but for character recognition, which is a different application to CBOD.}

{\color{black}
A. Ruospo et al. \cite{9217880} characterize the impact of software-injected permanent faults in the CNN weights of the LeNet-5 CNN \cite{726791} with the MNIST dataset \cite{deng2012mnist} for handwritten digit classification. Authors provide a characterization for 32-bit floating-point arithmetic and different fixed-point arithmetic configurations (32, 18, 16, 10, and 6 bits).
Similarly, 
F. Libano et al. \cite{9319148} study the impact of radiation induced hardware faults on different reduced precision models. In particular, the work analyses 32-bit and 16-bit floating-point versions, and an 8-bit integer version of a minimalist CNN (7 layers) also evaluated against the MNIST dataset \cite{deng2012mnist} for handwritten digit classification, which is a simpler task than CBOD.

Other works \cite{9833576,8854431} focus on the detection of faults exploiting spatio-temporal redundancy in the context of AD. Authors build on the fact that consecutive frames have a very high degree of visual similarity to detect faults. Authors compare the outcome of consecutive frames explicitly to detect hardware faults. Instead, we use the default implementation provided in the Darknet framework, which averages the confidence of the detections across the last three frames to mitigate some sporadic misdetections, but without including any specific mechanism for fault detection or correction. Instead, the very same mechanism used to mitigate model inaccuracies is exploited for fault tolerance in our case.
}

Overall, {\color{black}our paper is the first one providing} a complete assessment of different approximation domains applied incrementally in a real-life case study for AD systems taking into account the fault-tolerance that those systems provide due to time and space redundancies.

\section{Background}
\label{sec:back}

{\color{black}This section, provides background on the CBOD process, YOLO, and precision and approximation in arithmetic.}

{\color{black} \subsection{Camera-based Object Detection}}

\begin{figure}[t!]
	\includegraphics[width=1\columnwidth]{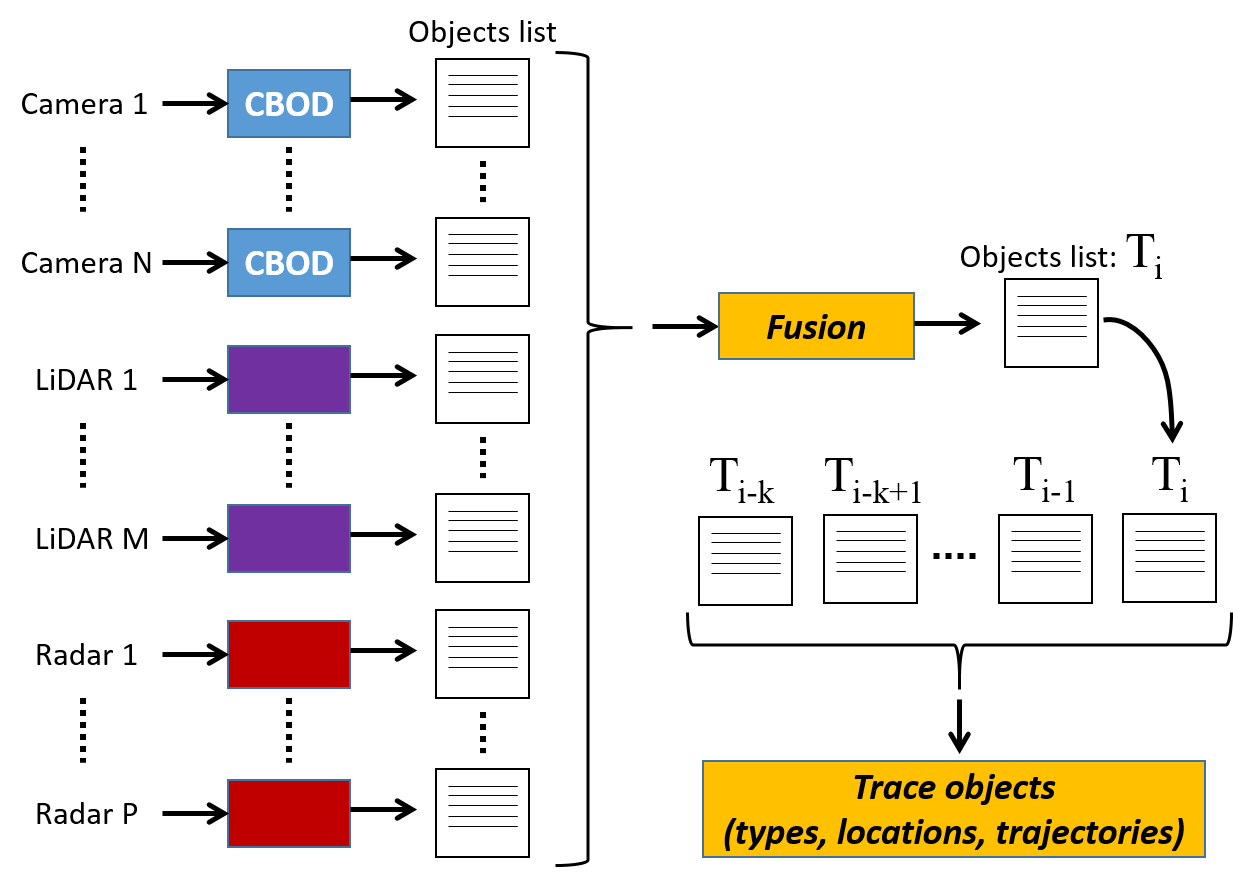}
	\caption{Perception process in AD.}
	\label{fig:percep}
\end{figure}

The perception process in AD frameworks, such as Apollo, builds upon complementary and redundant sensors for {\color{black}scene understanding}. This includes sensors of different types, such as cameras, LiDARs, and radars, but also sensors of the same type that may cover some surrounding areas redundantly (e.g. cameras with overlapping angles of view). This is illustrated in Figure~\ref{fig:percep}. 
Each sensor process generates a new object list periodically (e.g. every 40 ms for CBOD at a rate of 25 \emph{frames per second}, FPS), which are fused to generate a single object list. 

The fusion process needs to provide a coherent identification of the objects surrounding the car, which implies matching objects identified by different sensors, taking into account their type, size, location, and confidence in their detection. This process is intrinsically fault-tolerant since if one sensor discards a real object {\color{black}due to} low confidence in its detection, but the other sensors covering the same area detect it with high confidence, the object will be {\color{black}considered for} the fusion process. 

The fusion process periodically generates a new list of objects (e.g. at times $T_{i-2}$, $T_{i-1}$, $T_{i}$). Then, such sequences of lists of objects need to be processed to track objects over time so that their trajectories can be properly predicted, thus allowing the AD system to take safe driving decisions. The processing of consecutive object lists is, again, intrinsically fault-tolerant since detections are leveraged across lists over time, so sporadic false positives/negatives have no semantic impact {\color{black}on} the AD object detection process. {\color{black}As an illustrative example, if a car is detected in the front view of the vehicle (e.g. in object lists at times $T_{i-k}$, $T_{i-k+1}$,..., $T_{i-1}$) but suddenly it is not detected in one object list (e.g. $T_{i}$), the trajectory prediction module will still be capable of considering such car as long as the non-detection of that car occurs sporadically. Analogously, if a car is erroneously detected in front of the vehicle in just one object list (e.g. $T_{i}$), the trajectory prediction process will be capable of detecting it since it is impossible for such car getting in front of the vehicle in a too short time interval (e.g. in 40 ms), by being missing in $T_{i-3}$, $T_{i-2}$, and $T_{i-1}$, and instead, being in a specific location (e.g. close to the car) in $T_{i}$.}

Overall, the perception process of an AD framework is a fault-tolerant process where independent false positive/negative detections for one sensor, or jointly for multiple sensors instantly (e.g. in a single object list produced by the fusion process), do not have any semantic impact affecting the driving decisions.

In this work, we analyse {\color{black}several videos captured by a single camera in an autonomous vehicle to be processed by YOLO {\color{black}v3}, as CBOD representative. Hence, we do not exploit fault tolerance from redundant sensors {\color{black}due to lack of appropriate data for that purpose (e.g., data from multiple cameras from a driving sequence)}. However, we illustrate tolerance to errors of the process due to the intrinsic stochastic nature of the detection process in YOLO, and due to the fault tolerance of object detections across consecutive video frames.}

{\color{black}
\subsection{YOLO Real-time Object Detection System}
YOLO is a state-of-the-art real-time CBOD system used as the main component for CBOD in several industrial autonomous driving software frameworks \cite{apollo, apollo-perception, autoware, nvidia-drive}.
{\color{black}YOLO operates on floating-point 32-bit data, so we assume this number representation as the baseline case in the rest of the paper despite our findings could be applied analogously to other representations (e.g., integer numbers).}

\begin{figure}[t!]
	\includegraphics[width=1\columnwidth]{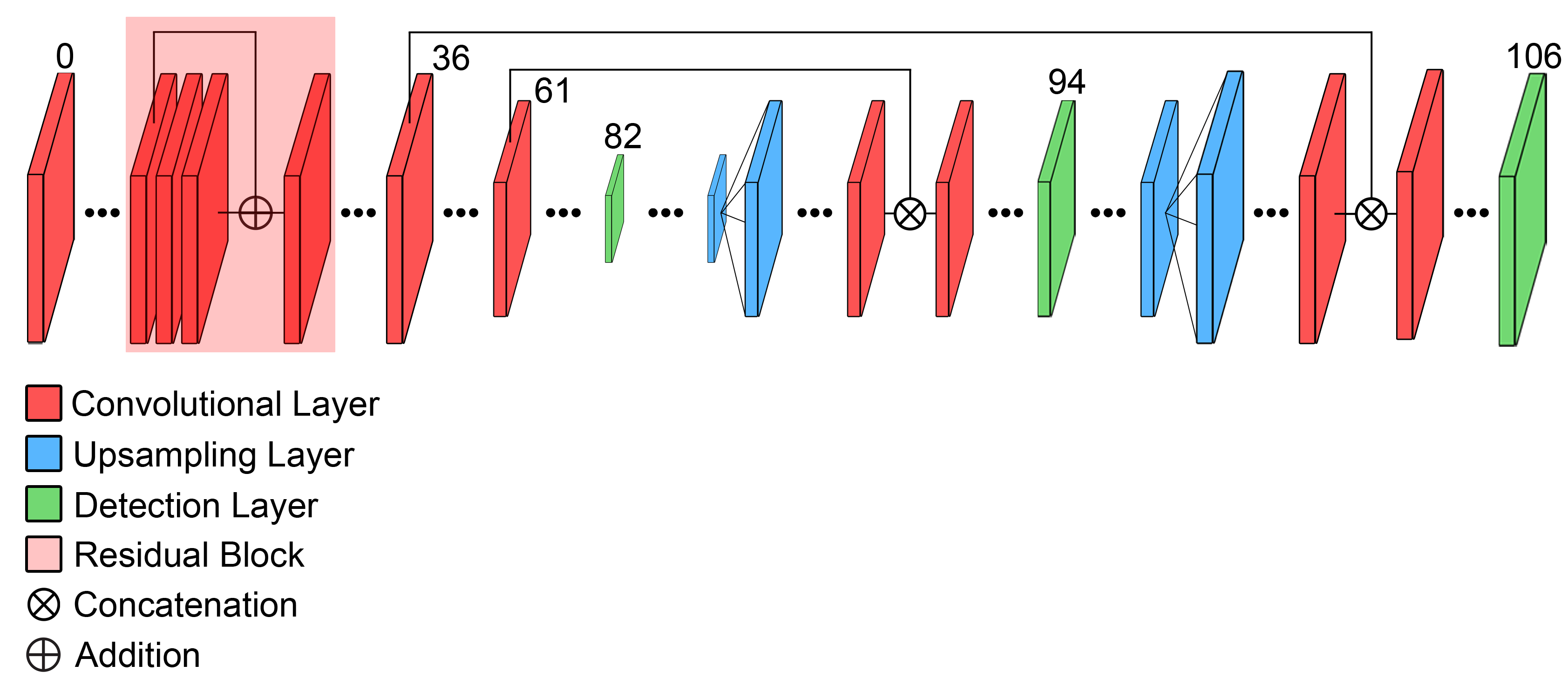}
	\caption{YOLO v3 architecture overview.}
	\label{fig:yolo-arch}
\end{figure}

\begin{figure}[t!]
\centering
\includegraphics[width=1\columnwidth]{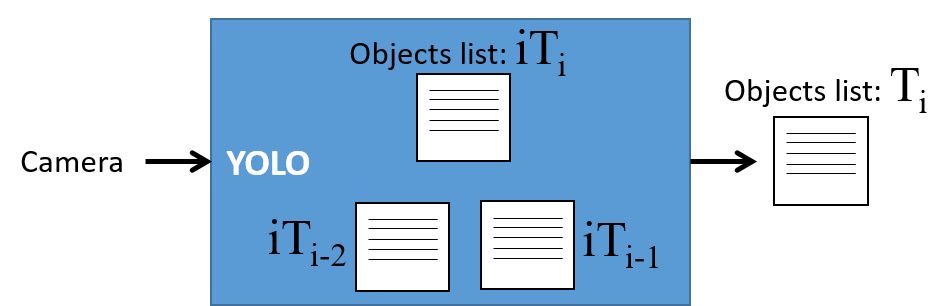}
\caption{YOLO case study evaluated.}
\label{fig:yolocasestudy}
\end{figure}

Figure~\ref{fig:yolo-arch} shows an overview of the architecture of YOLO v3. YOLO v3 consists of 106 layers that include Convolutional, Upsample, Shortcut, Route, and Detection layers. The Residual Blocks consist of several convolutional layers and skip connections that are used for feature extraction. The main unique feature of the architecture is that object detections are performed at three different scales. In particular, the detection processes are performed on layers 82, 94, and 106, and after each detection processes, the input image is upscaled by a factor of 2, allowing for better detection of objects of different sizes. The image is divided into a grid, and each cell is responsible for predicting three bounding boxes (i.e. a rectangle delimiting the area of an object). For each of the three bounding boxes, its coordinates, an objectness score (i.e. the probability that the cell includes an object) and class scores (i.e. a score for each class indicating the probability that the object belongs to that class) are calculated. This model defines 80 classes that include animals, food, vehicles, and pedestrians among others. Many bounding boxes may not contain an object, overlap with other bounding boxes that contain the same object, or have a very low probability of containing an object and/or belonging to a specific class. Therefore, a filtering process is applied to obtain the predictions with the highest confidence level and above a set threshold -- set to 50\% by default -- and discard the remaining predictions.} {\color{black} As illustrated in Figure~\ref{fig:yolocasestudy}, this can be done for the internal object list ($iT_i$) (i.e. an independent analysis of the frame), or at a higher abstraction level, for the resulting output list ($T_i$) that combines the detections of the last three frames in order to determine if a detection needs to be considered (i.e. the average confidence level is above the 50\% threshold).}

{\color{black}
\subsection{Precision and Approximation in Arithmetic}
\textbf{Reduced Precision Arithmetic.} Reduced precision arithmetic is a widely-adopted model compression approach \cite{10.1145/3079856.3080246, wu2020low, 7011421, gong2014compressing}. Reducing the number of bits employed for the number representation, by using for example {\color{black}floating-point 16-bit} or {\color{black}integer 16-bit}, allows the immediate reduction of the model size. The process to map values of a higher precision arithmetic into a lower precision arithmetic is called quantization. For instance, a given {\color{black}floating-point 32-bit} number that cannot be represented fully precisely with {\color{black}floating-point 16-bit} is \emph{quantized} by mapping it to a different -- yet very close -- {\color{black}floating-point 16-bit} number.
Reduced precision arithmetic can bring area, power, and timing savings due to the reduced hardware complexity.

\textbf{Approximate Arithmetic.} Approximate arithmetic \cite{9313992, 8617877, SurveyApproxALUs} consists of implementing simplified arithmetic units, such as adders and multipliers, whose results are less precise compared to fully-precise arithmetic with a similar number of bits. The advantage of approximate designs is that the hardware complexity is significantly reduced, which leads to area, power, and timing savings, at the cost of -- typically slightly -- reduced precision.

The aforementioned approaches may be combined together to obtain further model compression and energy improvements at the expense of potentially lower accuracy. 
}

\section{Approximation Domains and Impact}
\label{sec:analysis}

{\color{black}In this section, we analyse how the YOLO camera-based object detection in particular, and deep learning-based object detection in general, perform object detection to easily tolerate errors, we review the three approximation domains considered in this paper, and how these can affect the object detection outcome.}

\subsection{Impact of Approximation in YOLO}

YOLO {\color{black}v3} implements object detection with 100+ layers, out of which the most computing-intensive ones are 53 Convolutional layers devoting most of their time to matrix-matrix multiplications, which is at the heart of most Deep Learning frameworks~\cite{HamidISORC}. 
Inference occurs with the sequential processing of a number of matrices. Therefore, any inaccuracy or computing error occurring in the first layers is propagated to the following layers.

Because of the nature of the inference process, {\color{black}computationally speaking, contents of one cell of one matrix are used for the computation of multiple cells of another matrix which, in turn, is processed {\color{black}similarly}. Therefore, while} an error or an inaccuracy in the result of one cell is propagated to an increasing number of cells in the following layers of the network, such propagation occurs with decreasing intensity due to the use of weights, {\color{black}that lead only to partial propagation of errors to each individual cell.}
Overall, a local error in one cell becomes a series of much smaller errors in multiple cells in the following layers. Moreover, those errors can {\color{black}compensate the effect of each other partially}. {\color{black}For instance, a slightly overapproximated value (e.g. $0.5$ instead of $0.48$) added with a slightly underapproximated value (e.g. $0.4$ instead of $0.43$) may mitigate inaccuracies (e.g. $0.5+0.4=0.9$ instead of $0.48+0.43=0.91$).}
{\color{black}Similar compensation effects have already been observed in other domains such as, for instance, critical path delay of circuits where, the larger the number of gates in the path, the lower the variation due to statistical compensation~\cite{VariationsCompensation}.}

In semantic terms, such error propagation translates into some minor variations in the confidence levels and object locations for multiple objects detected, rather than creating large deviations affecting just one or very few objects. This is true when using lower precision arithmetic, approximate arithmetic, and even when having sporadic erroneous values.
Therefore, one could expect that only those detections whose confidence levels are close to the threshold of acceptance (e.g. 50\%) may change enough to lead to a different semantic output. For instance, 
if confidence levels change by up to $\pm$5\%, only detections with confidence levels originally in the range $[45\%,55\%]$ could be classified differently.

{\color{black}If we further consider that the fusion process leverages such information from different redundant or partially-redundant sensors\footnote{As indicated before, fault-tolerance due to the use of multiple redundant sensors is not analysed quantitatively in the use case in this paper.}, and then such information regarding object detections is also leveraged across periodic detection processes (e.g. every 40 ms), we can expect that objects in the critical confidence level range change across sensors, and over time for the same sensor. Moreover, we can also expect that inaccuracies do not always affect in the same direction a given object (e.g. moving it from slightly above the threshold to below the threshold, or vice versa, across several sensors and repeatedly over time).}

In summary, the stochastic nature of the deep learning-based object detection process, thus with estimated confidence values and thresholds, and with sensor and time redundancies, is intrinsically fault-tolerant and hence, amenable to the use of techniques that trade off some accuracy to save power and design complexity.

\subsection{Approximation Domains in Object Detection}

{\color{black}Our target for approximation are the arithmetic units based on the assumption that they generally account for most of the area and power in systolic arrays. Also, we build on the assumption that their latency determines, or at least has large impact, on the operating frequency of the accelerator.}

\textbf{Reduced precision arithmetic}. Reduced precision (e.g. 16-bit operands instead of 32-bit ones) brings several benefits such as the use of less area to store data and for the arithmetic units operating such data, lower latencies for arithmetic units operating those smaller operands, and lower power to perform computations among others. 
{\color{black} On the other hand, reduced precision can encode fewer values. Hence, whenever the value to be stored cannot be represented exactly with the number of bits available for the particular representation and precision used, a close value that can be represented is used instead. In general, the higher the precision, the higher the number of values that can be represented exactly, and the lower the error introduced due to rounding a non-representable value to a similar representable one.}

In the particular case of YOLO, {\color{black}similarly to many {\color{black}deep learning} models, the model parameters are represented with floating-point arithmetic}. By default, 32-bit floating-point IEEE754 compliant values (also referred to as floats) are used by YOLO, such that 1 bit is used to represent the sign ($s$), 8 bits for the exponent ($e$), and 23 bits for the mantissa ($m$). Instead, {\color{black}standard} 16-bit floating point values (also referred to as half-floats) devote 1 bit to $s$, 5 to $e$, and 10 to $m$. Other than denormalized values and other special cases, values represented have the form:
$$
half\_float(s,e,m) = (-1)^s \cdot 2^{e-15} \cdot (1.m_2)
$$
For instance, the half-float $0 01101 0101010100_2$ would correspond to the following real number:
\begin{eqnarray}
	& & half\_float(0,01101,0101010100) = \nonumber \\
	& & ~~~~~~~~~~ (-1)^0 \cdot 2^{13-15} \cdot 1.0101010100_2 = 0.333007813 \nonumber
\end{eqnarray}

Note that in the case of floats, values represented have the form $float(s,e,m) = (-1)^s \cdot 2^{e-127} \cdot (1.m_2)$.

In general, using $b$ bits for the representations implies that up to $2^b$ values can be represented. Therefore, by decreasing $b$, fewer values can be represented. Whether this has a large or small impact strictly depends on the values represented. For instance, if their exponents are typically in the range $[-14,15]$, then the number of $e$ bits of half-floats suffices, and {\color{black}all $m$ bits can be used for meaningful digits, as opposed to denormalized and out-of-range values,} thus limiting precision loss to be below 0.05\% (i.e. $\frac{1}{2048}$). In the case of YOLO, we note that most values are in this range. Moreover, values out of that range are normally added to values with higher exponents, thus diluting to some extent their impact in precision.

Overall, reduced precision arithmetic is expected to cause a limited impact in YOLO object detection if it is not reduced too aggressively (e.g. using only 8-bit floating-point representation).

\textbf{Approximate arithmetic}. There are a plethora of arithmetic unit designs for approximate arithmetic, such as approximate adders and multipliers. Those can be used to operate the mantissa of the operands, since exponents are much less tolerant to errors due to their exponential impact in the value represented, {\color{black}as shown in \cite{9217880,10.1145/3126908.3126964,8704548}}, and instead, the mantissa may tolerate errors particularly in its lower-order bits, which could cause effects similar to those of reduced precision arithmetic.

The rationale behind approximate arithmetic is that a significant part of the logic is devoted to providing precision for the computation of a reduced number of values. Thus, such logic can be removed or reduced so that the cost reduction is significant while the impact in the result is limited to some inputs, and in those cases, the impact is also low by using approximation for the lowest order bits only. 

As discussed for the case of reduced precision, YOLO uses values such that, in most cases, the highest order bits of the mantissa are relevant. Hence, the impact of approximation is very limited. Analogously to the case of reduced precision, {\color{black}inaccuracies due to approximate arithmetic impact the lowest order bits} thus mitigating the relative impact of those inaccuracies. 
{\color{black}And also, as in the case of reduced precision, errors due to approximate arithmetic can compensate each other partially.}

Overall, as for reduced precision arithmetic, approximate arithmetic is expected to cause a limited impact in YOLO object detection.

{\color{black}On the other hand, approximate arithmetic provides power reductions and a reduction of the critical path length (i.e., minimum affordable cycle time). For instance, for the approximate adder considered in our evaluation, authors show a 28\% power reduction and 50\% delay reduction~\cite{approxADD} (see further details in Section~\ref{sec:evaluation}). In the case of the multiplier, power reduction may be as high as 31\%~\cite{approxMUL}.}

\textbf{Aggressive low voltage operation}. Dynamic energy is the dominant source of energy consumption in combinational logic such as those arithmetic units used for compute-intensive workloads such as object detection. It is well-known that dynamic energy is quadratic on supply voltage ($V_{DD}$) and hence, one of the most effective ways to save energy in this type of units is decreasing $V_{DD}$, despite the negative impact that lower $V_{DD}$ operation can have on latency~\cite{Razor}.

Circuits are usually powered and timed with some guardbands to ensure that they can complete their function by the end of the cycle. However, given a pair \linebreak $<V_{DD},T_{cycle}>$, where $T_{cycle}$ stands for the cycle time, such that correct operation is ensured, one can decrease $V_{DD}$ while keeping $T_{cycle}$ so that lower energy is consumed at the expense of some sporadic timing errors that will lead to erroneous outputs. Such errors can occur for some specific inputs triggering specific delay paths in the circuits, affected negatively by process variations, under some specific temperature ranges, or under particular combinations of those effects. 
{\color{black}Alternatively, a simpler approach consists of using lower power -- yet slower -- transistors without changing $V_{DD}$ (e.g. synthesizing those parts of the circuit targeting a slightly lower operating frequency while keeping the original frequency during operation), so that power savings can be obtained while keeping $V_{DD}$ unchanged. Slower paths due to slower transistors may lead to some errors under specific operation conditions and input data transitions, with the impact of errors and power savings being completely implementation dependent.}
The impact of such timing errors is generally arbitrary because typically there are many critical paths across several corner conditions with a delay close to $T_{cycle}$.

Overall, decreasing $V_{DD}$ aggressively brings significant dynamic energy savings at the expense of causing some sporadic errors with arbitrary impact in the output result of arithmetic units. However, if those errors are rare enough, the overall impact in the object detection process is expected to be tiny in semantic terms.

{\color{black}

\section{Evaluation Framework}
\label{sec:framework}

In this section, we present our evaluation framework, discussing the datasets used for the evaluation, the approximation setups evaluated, and the precision metrics employed to assess the accuracy of the model.
}

\subsection{Datasets}
\label{sec:datasets}

To assess the impact of different approximation domains in the CBOD process, we consider {\color{black}the default YOLO v3~\cite{YOLOv3} version with a pre-trained set of weights that is implemented within the Darknet framework~\cite{darknet},} processing a set of images with labelled objects -- the COCO dataset~\cite{cocodataset} --  {\color{black}as well as several representative driving video recordings~\cite{VideoYOLO, video-1, video-2, video-3, video-4, video-6}.}

{\color{black}The labelled dataset consists of a set of images that have been tagged with labels accurately. In the case of object detection, a bounding box and an associated class are set for every object contained in each image. This information is referred to as the ground truth and is later used for comparison against the output of the CBOD process. We use the {\color{black}testing subset of the} COCO dataset for our case study, as this dataset contains thousands of {\color{black} high-quality} images. {\color{black}We can expect relatively highly accurate results with this dataset, since we use a pre-trained version of YOLO that was trained using the training subset of COCO.} The COCO dataset contains images of common objects in context. However, due to its generality, the dataset includes many images completely unrelated to driving scenarios. Therefore, we filtered out those images not containing either a person or a vehicle to reduce the number of unrelated images. However, even if the images contain vehicles, this does not imply that the images strictly belong to driving scenarios.

To complement our experimental analysis, we have studied and included a set of unlabelled videos recorded by collection cars explicitly for the development of CBOD. Due to the cost of performing object detection with software-implemented arithmetic units (i.e. 45 minutes per frame and configuration in our case), we only processed a subset of the frames in each video ($\approx$ 30 seconds for each video).}

{\color{black}
We considered the use of several autonomous driving labelled datasets, such as, CityPersons \cite{citypersons}, KITTI \cite{kitti} and ApolloScape \cite{apolloscape}, among others (up to 12 datasets in practice). However, to the best of our knowledge, there are no publicly available datasets for which to evaluate the pre-trained YOLO v3 model with sufficient accuracy.
We discarded each of those datasets due to at least one of the following reasons: (i) not being publicly available, (ii) not being sufficiently representative for our case study, due to not having the most relevant object classes found in the context of autonomous driving (i.e., at least vehicles and pedestrians), (iii) missing 2D annotations, which are required since the pre-trained YOLO v3 model produces 2D bounding box object detections, and (iv) using similar annotation guidelines to those of the pre-trained YOLO v3 model, which is trained with the COCO dataset, to guarantee meaningful results.
}

\subsection{Approximation setups} 

We have integrated the SoftFloat library~\cite{softfloat} in {\color{black}the Darknet framework} to implement {\color{black}floating-point numbers at software-level, replacing native floating-point operations by their software emulation, which allows us to use arbitrary representations and arithmetics. We evaluate four} different scenarios: the IEEE754 {\color{black}floating-point 32-bit} fully precise version (YOLO32full), the IEEE754 {\color{black}floating-point 16-bit} fully precise version (YOLO16full), the {\color{black}floating-point 16-bit} version using approximate multipliers and adders (YOLO16approx), and YOLO16approx with some faults injected to mimic erroneous outputs due to aggressive low voltage operation (YOLO16appfault). 

{\color{black}As said, YOLO16approx corresponds to YOLO16full, but using approximate multipliers and adders for the mantissa, {\color{black}which have been implemented on top of the SoftFloat library to emulate their behavior}. In particular, YOLO16approx implements the addition operation without forwarding the carry from the lowest half of the addition to the highest half of the addition~\cite{approxADD}. This optimization significantly decreases the latency of the most critical delay path, as shown in \cite{approxADD}, which enables the use of lower power gates within a given $V_{DD}$ and $T_{cycle}$ envelope, while still preserving full precision for the addition of the most significant bits, hence limiting the impact of approximation. Regarding the approximate multiplier, we use the design in \cite{approxMUL}, which provides a method to build 2x2 approximate multipliers and also provides a method to build larger multipliers using blocks of 2x2 smaller multipliers. Using this method, given 2 operands (A and B), split into their half uppermost bits (AH and BH respectively) and their half lowermost bits (AL and BL respectively), we have implemented an approximate multiplier where the blocks ALxBL, AHxBL and ALxBH are approximate, and the block AHxBH is accurate since this block has a higher impact in the result of the multiplication.}
{\color{black}While other approximate arithmetic units could be considered, the units considered prove to be effective, as shown later in this section, and hence, they already serve the purpose of illustrating that the use of approximate arithmetic is suitable path for CBOD in AD systems.}

Finally, YOLO16appfault implements YOLO16approx, but mimicking the impact of faults due to low voltage operation {\color{black}at software level, inside the SoftFloat library.} {\color{black}We model low voltage related faults} by randomly flipping one bit of the result of {\color{black}floating-point} multiplications or additions with a given probability {\color{black}of fault injection per result,} $p_{faulty}${\color{black}, which suffices to corrupt the result on an arbitrary manner}. {\color{black}How specific $p_{faulty}$ values relate to particular $V_{DD}$ values is technology dependent and beyond the scope of this work.}
Bits are only flipped in the mantissa. Given that the exponent and sign have fewer bits, {\color{black}we assume that the logic to operate them has lower delay (e.g. in terms of fanout-of-4, FO4, gates)} and can easily {\color{black}tolerate some delay increase to deliver correct results despite low voltage operation without impacting cycle time. Instead, we assume that mantissa bits are in the critical path for delay, and hence, any increase in the latency of the logic operating mantissa bits due to lower voltage operation could lead to erroneous outputs. Therefore, it is possible decreasing $V_{DD}$ to some extent with no impact in the exponent and sign bits, and experiencing faults only in the mantissa bits. Moreover, as shown later, mantissa bits are relatively fault tolerant, and hence, there are opportunities to trade off some faults in the mantissa for power savings.}
We evaluate some $p_{faulty}$ values to assess the sensitivity of different tradeoffs between $V_{DD}$ values {\color{black}(or degrees of aggressiveness using lower power transistors for the mantissa operation)} and fault rates generated by raising $p_{faulty}$ until it had a visible -- yet affordable -- impact in results.
{\color{black}Note that, processing each frame involves operating in the order of $10^{10}$ 16-bit floating-point values, so $10^{11}$ mantissa bits (the mantissa has exactly 10 bits for 16-bit floating point numbers). Hence, the number of faults injected is around $10^{10} * p_{faulty}$ per frame. Therefore, we injected between $10^4$ ($p_{faulty} = 10^{-6}$) and $10^7$ ($p_{faulty} = 10^{-3}$) faults per frame.}

\subsection{Precision metrics}

{\color{black}
To assess the impact on the accuracy of the different approximation domains, we} build on the Intersection over Union ($IoU$), Average Precision ($AP$) {\color{black}and Mean Average Precision ($mAP$)} metrics as described in \cite{APmetric}. The $IoU$ is obtained as follows:
\begin{equation}
	IoU = \frac{Area_{gt} \cap Area_{p}}{Area_{gt} \cup Area_{p}}
\end{equation}
where $Area_{gt}$ is the area of the ground truth bounding box for the corresponding object and $Area_{p}$ is the area of its predicted bounding box.
The $IoU$ gives a value between 0 (if the bounding boxes do not intersect) and 1 (if the bounding boxes are exactly the same in terms of area and position). Using $IoU$ we can set a threshold $t$ such that only if $IoU \geq t$, the prediction is regarded as correct, otherwise it is regarded as incorrect. In our case, we assume $t=0.5${\color{black}, in line with the PASCAL VOC challenge~\cite{PASCALVOC}}.  
With $IoU$ we can classify objects as correct detections (true positives, TP), erroneous detections (false positives, FP), and missed detections (false negatives, FN).

{\color{black}We have used a publicly available framework \cite{APmetric} to measure the mAP and to obtain the number of TP, FP, and FN of the different experimental setups.}

{\color{black}
The assessment of object detectors is mainly based on the \emph{Precision} ($P$), an indication of the accuracy for detecting only relevant objects (i.e., objects represented in the ground truth), and \emph{Recall} ($R$), an indication of the accuracy for detecting all relevant objects. Precision and recall are defined as follows:

\begin{align} \label{eq:precision-recall}
	\begin{split} 
		P &= \frac{TP}{All\, detections} = \frac{TP}{TP + FP}
	\end{split}\\
	\begin{split}
		R &= \frac{TP}{All\, ground truths} = \frac{TP}{TP + FN}
	\end{split}
\end{align}

A good detector should identify all ground-truth objects and avoid predicting non-existent objects. Therefore, FP and FN should be close to 0, and $P$ and $R$ close to 1.

Building on these two concepts, the Average Precision ($AP$) is measured by an all-point interpolation method for a given object class, and a set $IoU$. The $AP$ is defined as follows:
\begin{equation} \label{eq:ap}
	AP_{0.5} = \sum_{n} (R_{n+1} - R_n)P_{interp}(R_{n+1}) 
\end{equation}
where $AP_{0.5}$ indicates that the $AP$ is measured with $IoU$ = $0.5$ and
\begin{equation}
	P_{interp}(R_{n+1}) = \max_{\{\tilde{R}:\tilde{R} \geq R_n + 1\}} P(\tilde{R}) 
\end{equation}

If $AP$ values are obtained for different object classes, they can be averaged to obtain a single value called Mean Average Precision ($mAP$) that weights all object classes equally disregarding the object frequency for each type~\cite{APmetric}. The $mAP$ is defined as follows:
\begin{equation}\label{eq:map}
	mAP = \frac {1}{N} \sum_{i=1}^N AP_i 
\end{equation}

}

{\color{black}

\section{Results}
\label{sec:evaluation}

In this section, we discuss the labelled dataset evaluation, the video case study evaluation, the confidence range of the detections, and we provide some complementary experiments analysing the area of the misdetections and the misclassification of vehicle types.

}

\subsection{Labelled Dataset Results}

We have computed the $AP$ values for the labelled dataset for all types of vehicles (cars, trucks, motorbikes, and buses) and for the person class of object. Results are shown in Table~\ref{tb:mAP_video_0} (top data rows), where we show the number of TP{\color{black}, FP and FN}, the $mAP$ for the four vehicle classes, and for the four vehicle classes and persons together. In particular, we show results for YOLO32full, YOLO16full, YOLO16approx, and YOLO16appfault varying $p_{faulty}$ between $10^{-6}$ and $10^{-3}$. As shown, reducing precision slightly reduces both the number of TP and FP, and slightly increases FN, {\color{black}in this case,} which still leads to near-identical $mAP$ values to those of YOLO32full. If we further introduce approximate arithmetic (YOLO16approx), results remain nearly constant. If we introduce faults, results change marginally up to $p_{faulty}=10^{-4}$, and higher fault rates lead to noticeable differences due to the increased number of FP.

Overall, these results indicate that CBOD is highly tolerant to multi-domain approximations such as those of reduced precision, approximate arithmetic, and even some moderate fault rates due to aggressive low voltage operation.

\subsection{Case Study (Video) Results}

\begin{figure*}[t!]
\centering
\includegraphics[width=0.32\columnwidth]{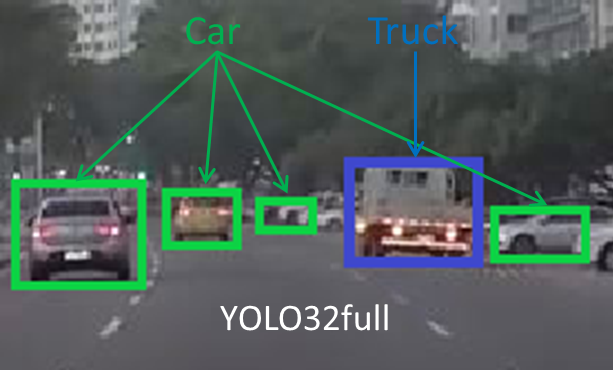}  
\vspace{0.2cm} \includegraphics[width=0.32\columnwidth]{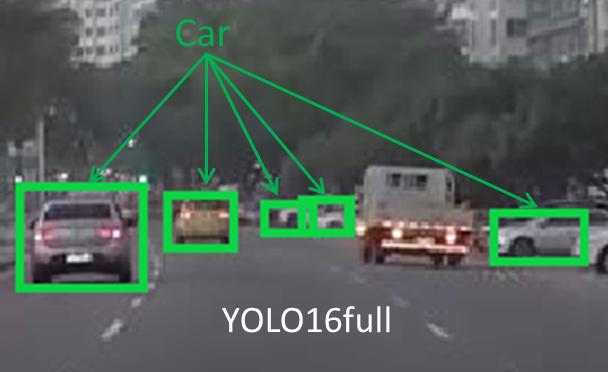} 
\vspace{0.2cm} \includegraphics[width=0.32\columnwidth]{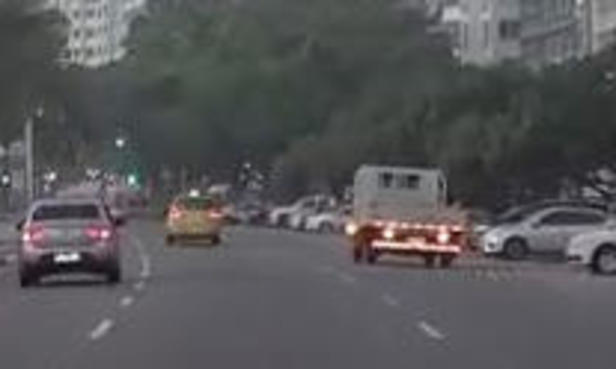}
\caption{Example of a frame with different (but true) object detections by YOLO32full (left), YOLO16full (center) and the original frame (right).}
\label{fig:pictures}
\end{figure*}

{\color{black}The main drawback of the labelled dataset analysis is that the COCO dataset does not contain many images relevant for driving conditions (e.g. images of road scenes taken by in-vehicle cameras), so the value of the results of this analysis in the context of autonomous driving are limited. 

We include the analysis of videos of real driving conditions that contain relevant data despite not being labelled{\color{black}, as discussed in Section~\ref{sec:datasets}}. The main drawback of using unlabelled videos is that they lack groundtruth labelled objects to assess the accuracy of the model. Therefore, the only means of assessing the accuracy of the model is to use the baseline YOLO32full as the reference configuration. However, the baseline configuration is not 100\% accurate. Therefore, when comparing different configurations, it cannot be stated whether discrepancies among them are due to erroneous detections in one or the other configuration, other than by visual inspection of the corresponding frame.} For instance, this is illustrated in the frame in Figure~\ref{fig:pictures}, where we see that both, YOLO32full and YOLO16full, have 5 TP and at least 1 FN. However, without visual inspection and taking YOLO32full as a reference, we would assume that YOLO32full has 5 TP and no FN, whereas YOLO16full has 4 TP, 1 FN, and 1 FP. {\color{black}It is important to clarify that all the quantitative results provided in this work are calculated strictly from the output of YOLO, and the metrics have been obtained with a scientific framework to avoid any kind of subjective visual inspection. Therefore, visual inspection is only employed for qualitative discussions.} {\color{black}This may lead to pessimistic accuracy results, to some extent. However, we believe that the analysis of unlabelled videos is still very relevant in this context because they allow considering real driving conditions.}

\begin{table}[t!]
\caption{Results for the labelled data set (top), and both $iT_i$ and $T_i$ for the video use case. {\color{black}Gray background rows indicate the most aggressive configurations in terms of approximation that cause negligible (below 1\%) accuracy degradation.}}
\label{tb:mAP_video_0}
\setlength{\tabcolsep}{1.0mm} 
\centering
\begin{tabular}{|l||c|c|c||c|c|}
\hline
\multicolumn{6}{|c|}{\textbf{Labelled data set}} \\
\hline
\textbf{Configuration} & TP & FP & FN & \multirow{2}{15mm}{mAP vehicle} & \multirow{2}{24.1mm}{mAP vehicle and person} \\
                       &    &    &    &             & \\
\hline
\textbf{YOLO32full}                & 5269 & 1076 & 4135 & 59.73 & 60.47\\ 
\hline
\textbf{YOLO16full}                & 5226 & 1038 & 4178 & 59.79 & 60.39\\ 
\hline
\textbf{YOLO16approx}              & 5230 & 1023 & 4174 & 59.49 & 60.16\\ 
\hline
\textbf{YOLO16appfault($10^{-6}$)} & 5238 & 1030 & 4166 & 59.26 & 60.00 \\ 
\hline
\rowcolor[gray]{.8}\textbf{YOLO16appfault($10^{-5}$)} & 5214 & 1018 & 4190 & 59.29 & 59.99\\ 
\hline
\textbf{YOLO16appfault($10^{-4}$)} & 5227 & 1034 & 4177 & 59.20 & 59.95\\ 
\hline
\textbf{YOLO16appfault($10^{-3}$)} & 5210 & 1219 & 4194 & 58.53 & 59.39\\ 
\hline
\multicolumn{6}{|c|}{\textbf{Video use case: $iT_i$ -- single frame detections}} \\
\hline
\textbf{Configuration} & TP & FP & FN & \multirow{2}{15mm}{mAP vehicle} & \multirow{2}{24.1mm}{mAP vehicle and person} \\
                       &    &    &    &             & \\
\hline
\textbf{YOLO16full}                             & 51297 & 857 & 1472 & 94.78 & 95.00 \\
\hline
\textbf{YOLO16approx}                       & 51142 & 797 & 1627 & 93.69 & 93.96 \\
\hline
\textbf{YOLO16appfault($10^{-6}$)} & 51123 & 842 & 1646 & 95,50 & 95,50 \\
\hline
\rowcolor[gray]{.8}\textbf{YOLO16appfault($10^{-5}$)} & 51061 & 839 & 1708 & 94,07 & 94,19 \\
\hline
\textbf{YOLO16appfault($10^{-4}$)} & 50804 & 1223 & 1965 & 90,71 & 91,21\\
\hline
\textbf{YOLO16appfault($10^{-3}$)} & 48853 & 3562 & 3916 &  83,14	 & 83,77\\
\hline
\multicolumn{6}{|c|}{\textbf{Video use case: $T_i$ -- confidence averaged across frames}} \\
\hline
\textbf{Configuration} & TP & FP & FN & \multirow{2}{15mm}{mAP vehicle} & \multirow{2}{24.1mm}{mAP vehicle and person} \\
                       &    &    &    &             & \\
\hline
\textbf{YOLO16full}                             & 49720 & 462 & 1184 & 96.25 & 96.21\\           
\hline
\textbf{YOLO16approx}                       & 49552 & 395 & 1352 & 97.29 & 97.14 \\
\hline
\textbf{YOLO16appfault($10^{-6}$)} & 49514 & 408 & 1390 & 97,06 & 96,92  \\
\hline
\rowcolor[gray]{.8}\textbf{YOLO16appfault($10^{-5}$)} & 49448 & 446 & 1456 & 95,42 & 95,62  \\
\hline
\textbf{YOLO16appfault($10^{-4}$)} & 49307 & 710 & 1597 & 92,27 & 92,84  \\
\hline
\textbf{YOLO16appfault($10^{-3}$)} & 47659 & 2422 & 3245 & 89,35 & 88,88  \\
\hline
\end{tabular}
\end{table}

Table~\ref{tb:mAP_video_0} (center and bottom rows) shows analogous results to those of the labelled dataset but for the unlabelled video. {\color{black} In the table, $iT_i$ corresponds to the object detection per individual frame (internal YOLO results), whereas $T_i$ corresponds to the object detection results averaged across the last three frames, as illustrated in Figure 3.}

Note that YOLO32full is omitted in the table since it is used as a reference to compare the other setups. Results are shown for both, the internal object list for a single frame (central set of rows) and for the list averaging confidence levels across 3 consecutive frames (bottom set of rows). The first observation is that $mAP$ values are very high for all configurations and object types (i.e. vehicles and persons) with respect to YOLO32full, except when $p_{faulty} \ge 10^{-4}$. This holds for both $iT_i$ and $T_i$ object lists. 

When comparing results for both lists, we observe that $T_i$ has lower TP{\color{black}, FP and FN} values. Lower TP, rather than indicating a higher number of undetected objects, it indicates that even \linebreak YOLO32full detects fewer objects because the intrinsic fault tolerance of YOLO averaging confidence levels across frames allows removing sporadic marginal detections (e.g. objects whose confidence is slightly above 50\% only sporadically). This very same effect also occurs for the other configurations. FP {\color{black}and FN} values also decrease due to the filtering of sporadically different confidence values for all configurations. Overall, globally, $mAP$ values for $iT_i$ and $T_i$ do not differ much. However, if we look at how fault-tolerant are both lists, we realize that $T_i$ is much more tolerant to approximation since, as we introduce further approximation domains and as we increase $p_{faulty}$, $mAP$ degrades slowly. For instance, the two last configurations for $iT_i$ bring $mAP$ drops of around {\color{black}3}\% and {\color{black}7}\% respectively, whereas for $T_i$ those drops are in the order of 3-4\% in both cases. 

{\color{black}We also observe that the impact of approximation is analogous across object types, as expected, since those are orthogonal concerns. Instead, a relationship exists between the dimensions of the objects{\color{black}, occupied area,} and the impact of approximation, since detections of small objects tend to have lower confidence values and the impact of approximation could make those values drop below or grow above the acceptance threshold, hence with a semantic impact.}

\begin{figure*}[t!]
\centering
\subfloat[YOLO16full]{\includegraphics[width=.49\linewidth]{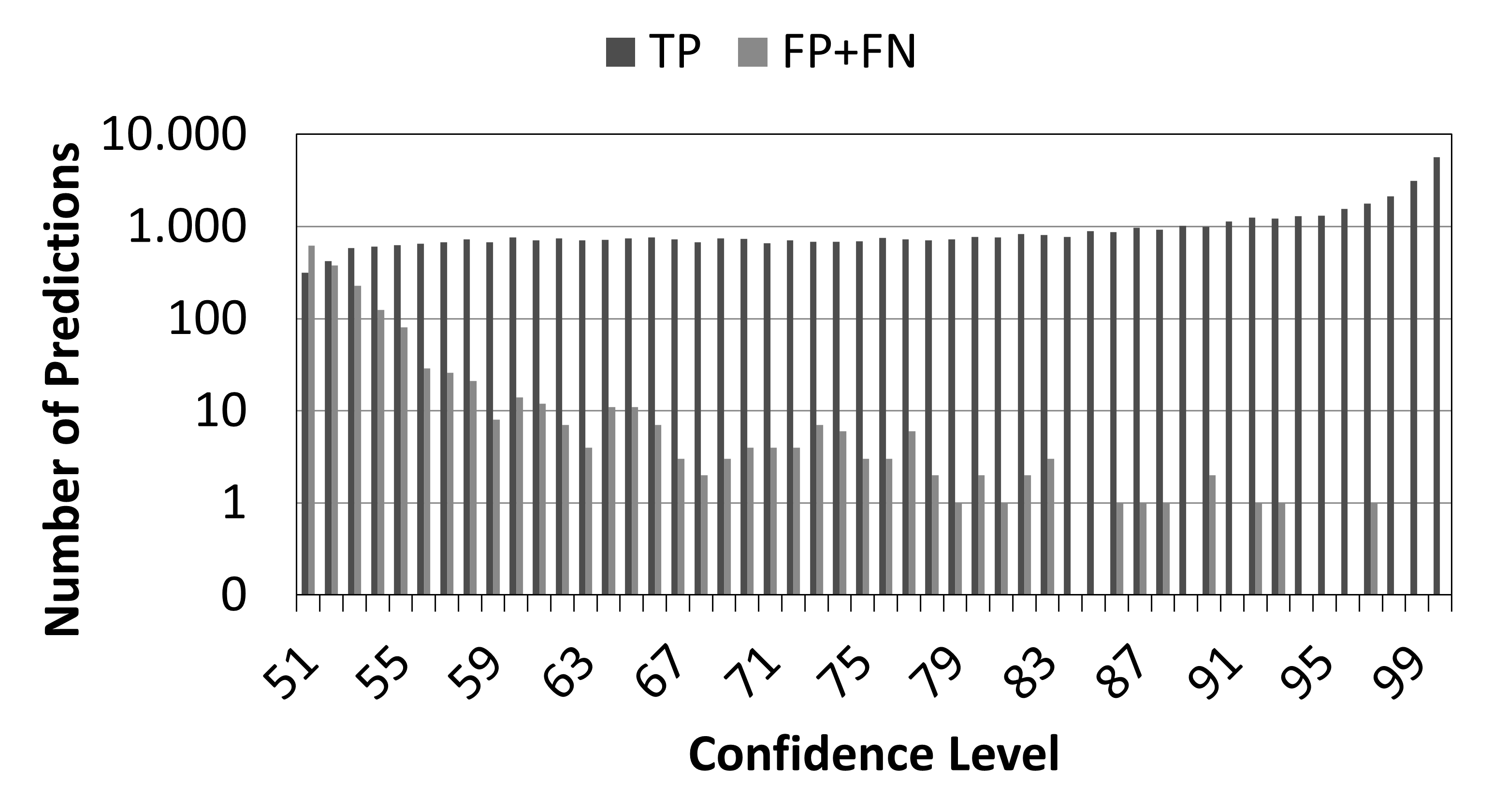}} \hfill
\subfloat[YOLO16approx]{\includegraphics[width=.49\linewidth]{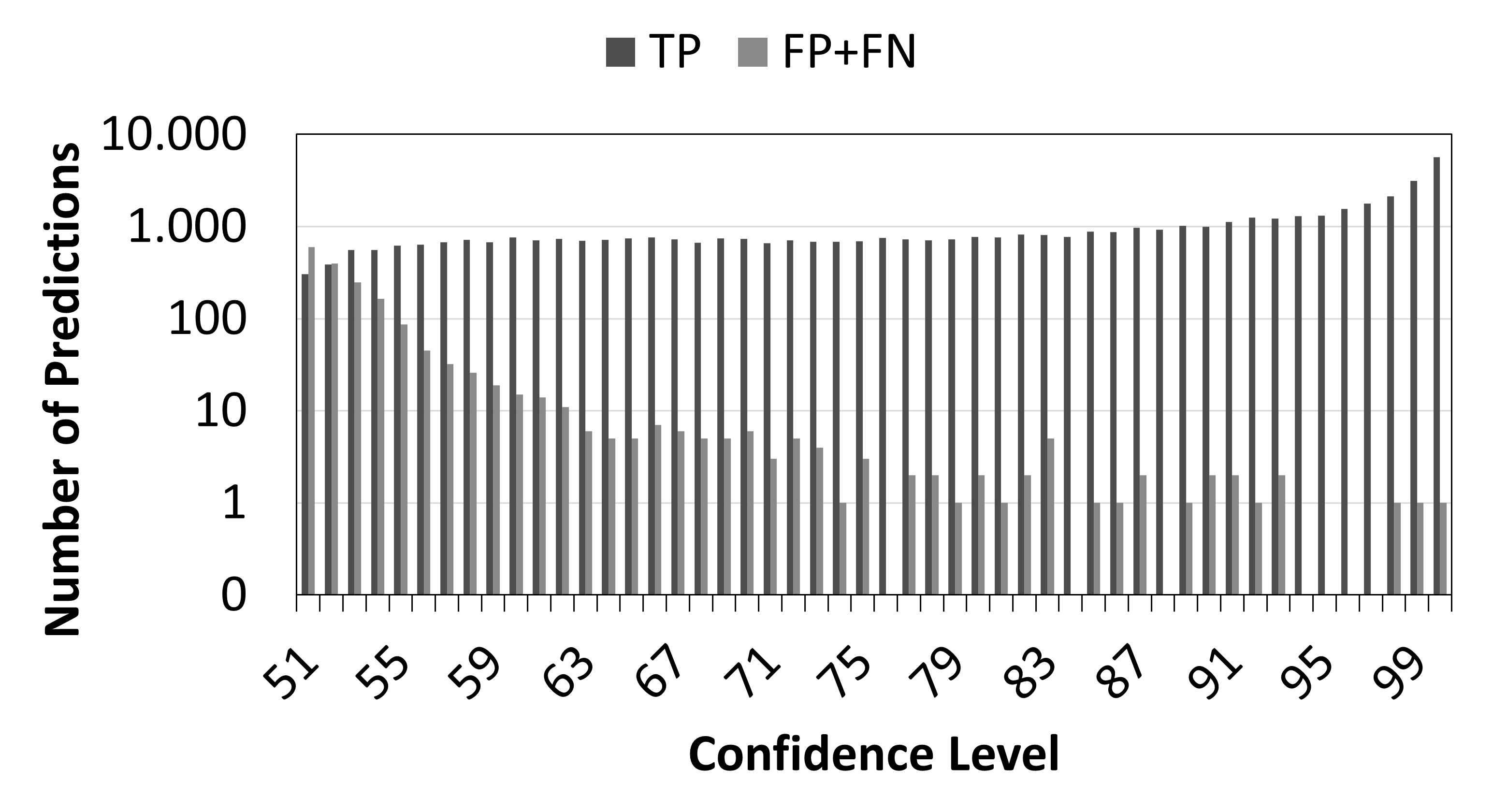}} \par
\subfloat[YOLO16appfault]{\includegraphics[width=.49\linewidth]{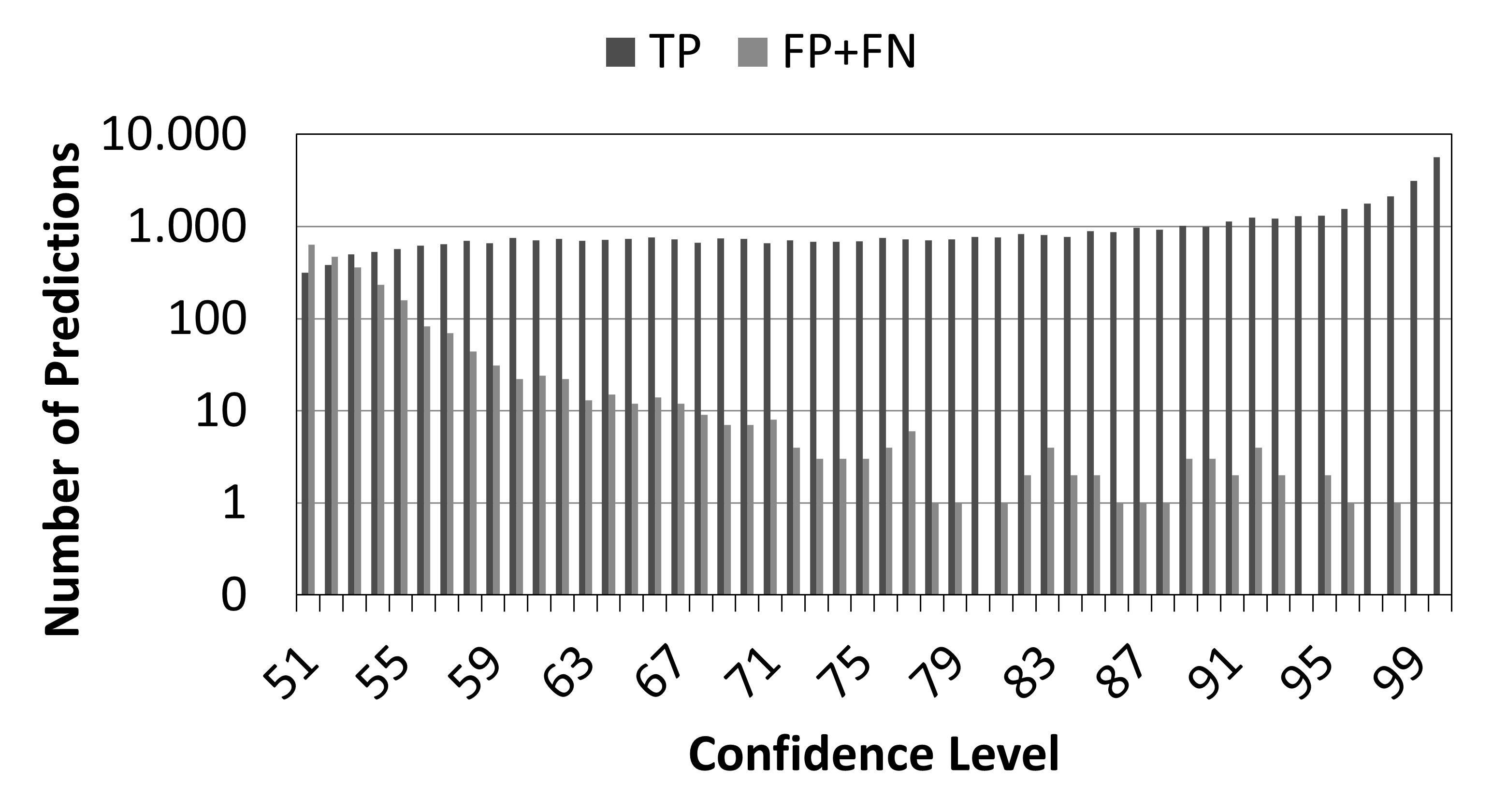}}
\caption{{\color{black}TP and FP+FN per confidence level for all object types with the $T_i$ object list. The configurations are (a) YOLO16full, (b) YOLO16approx, and (c) YOLO16appfault($10^{-4}$).}}
\label{fig:confidencelevels}
\end{figure*}

Finally, considering only whether predictions are above or below the threshold omits information about the confidence of those predictions. For this reason, we have measured the distribution of the confidence levels for both TP and {\color{black}FP+FN}. We depict those values for all object types (not only vehicles and persons) in Figure~\ref{fig:confidencelevels}, {\color{black}for the YOLO16full, YOLO16approx, and YOLO16appfault($10^{-4}$) configurations, using the $T_i$ object list.} Note that the y-axis is in logarithmic scale.

{\color{black}All configurations show similar trends.} As shown in {\color{black}Figure~\ref{fig:confidencelevels}}, TP distributes quite homogeneously across confidence levels. However, {\color{black}FP+FN} concentrates in the low confidence range, thus indicating that discrepancies negligibly impact objects detected with high confidence. This indicates that the object detection process is naturally fault-tolerant in the context of autonomous driving. 
Moreover, visual inspection revealed that discrepancies for high-confidence objects relate to (1) objects correctly identified in terms of location and type, but whose bounding box differs noticeably because other objects have been included in the bounding box, either in the approximate configuration or in the reference one (YOLO32full), and (2) overlapping objects where each configuration detects a different subset.

Overall, despite only a subset of redundancy levels are exploited in this work, we already show that CBOD for automotive systems is an intrinsically fault-tolerant process, which eases the adoption of specific accelerators building on multiple approximation domains such as reduced precision, approximate arithmetic, and aggressive voltage scaling for low power operation. 
{\color{black}In this work we have analysed the different sources of approximation in an incremental manner. However, we believe that the conclusions obtained with our analysis can be extrapolated to the other cases. 
For instance, reduced precision arithmetic and approximate arithmetic have some impact on the lowermost values of the mantissa, and rarely propagate to uppermost values. Hence, those effects are expected to hold regardless of the baseline used in each case. Similarly, fault injection in the mantissa has low impact when occurring in lowermost bits. If applied on YOLO32full, it is expected that a fraction of faults will impact the 10 uppermost bits of the mantissa (i.e., 10 every 23 faults on average), hence with similar effects to those of fault injection in YOLO16full, whereas the rest of the faults (i.e., 13 every 23 faults on average) will impact the 13 lowermost bits of the mantissa, hence with lower impact than that of injecting faults in the 10th bit of the mantissa of YOLO16full.
}

\subsection{Complementary Experiments and Results}

\subsubsection{Area of the Misdetections}

Another factor to take into account is the area of the misdetected objects of certain types. For instance, a car with a very small bounding box area corresponds to a car that is far away. Hence, it is not problematic if this car is misdetected for a few frames as long as the car is detected when the distance is reduced and the area of the bounding box of the car is much larger.

\begin{figure}[t!]
	\centering
	\includegraphics[width=1.0\columnwidth]{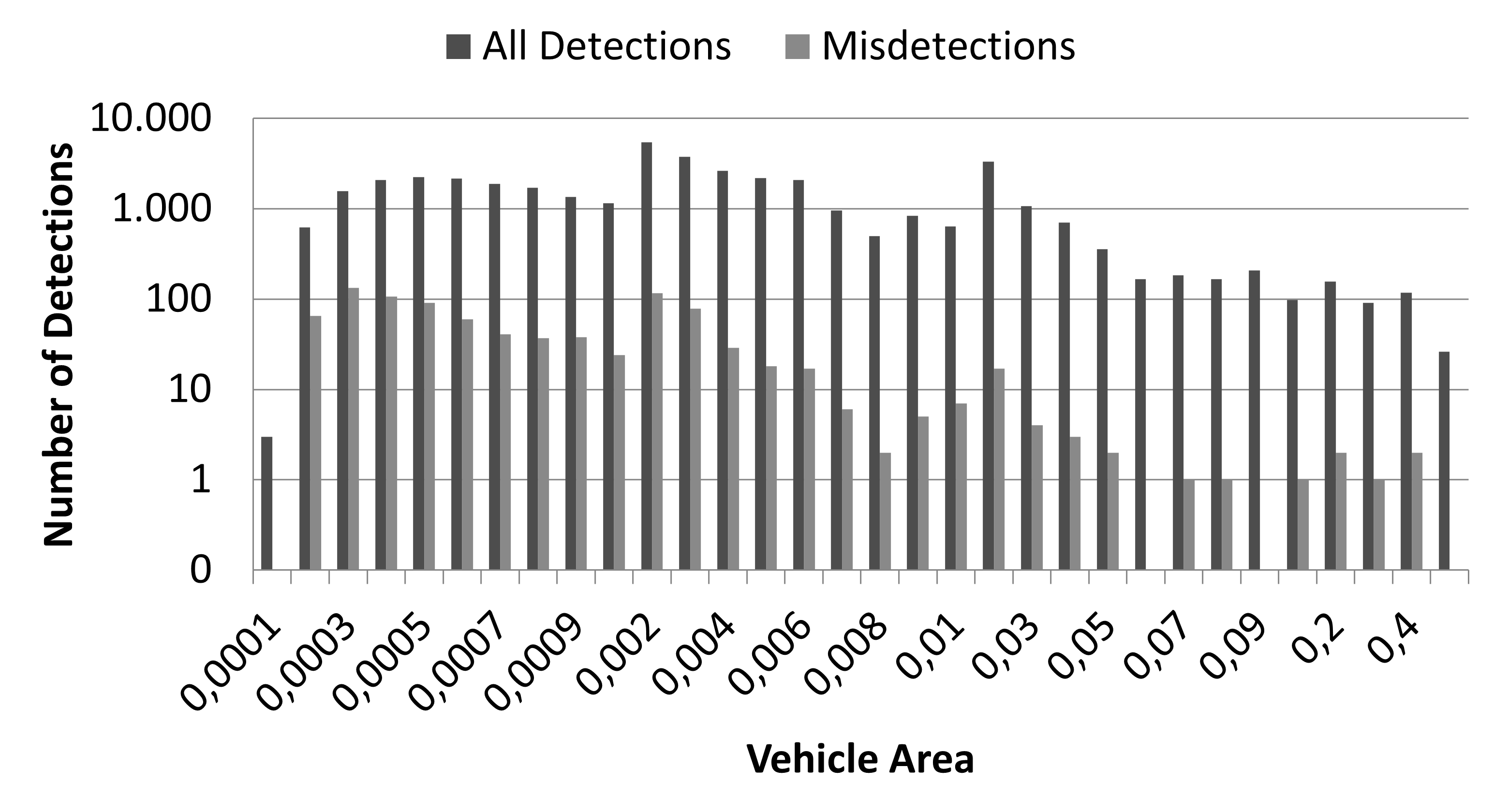}
	\caption{Area of {\color{black}all the objects detected and misdetected} for the YOLO16Approx $T_i$ video use case.}
	\label{fig:area-misdet-yolo16app}
\end{figure}

Figure~\ref{fig:area-misdet-yolo16app} shows the area range of all vehicle detections (TP) and objects not detected (FN) for the YOLO16approx $T_i$ video use case. The area of the objects is defined as the ratio of the bounding box area with respect to the total area of the image, hence the value of the area ranges from 0 to 1 and, for example, a 0.5 area corresponds to an object occupying half of the image. Note that the y-axis is in logarithmic scale. Note also that the x-axis increments in orders of magnitude every 10 steps.

\begin{figure*}[t!]
\centering
\includegraphics[width=0.48\columnwidth]{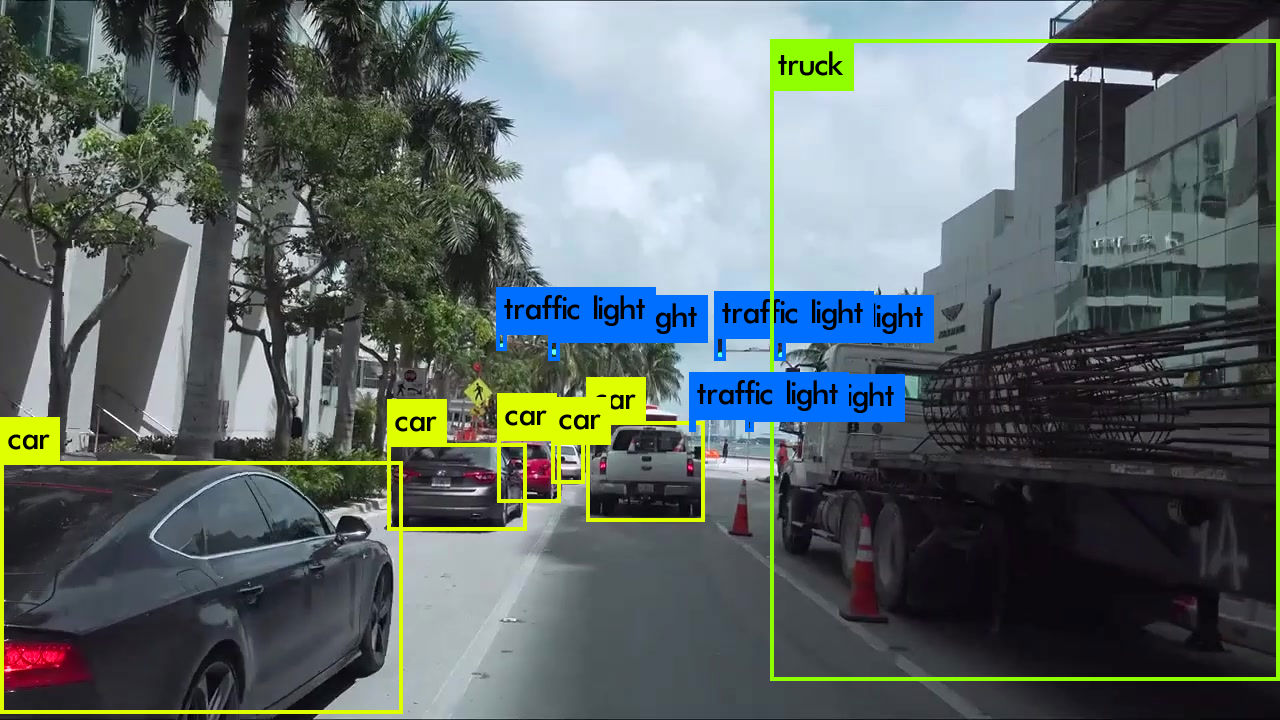}
\vspace{0.2cm} \includegraphics[width=0.48\columnwidth]{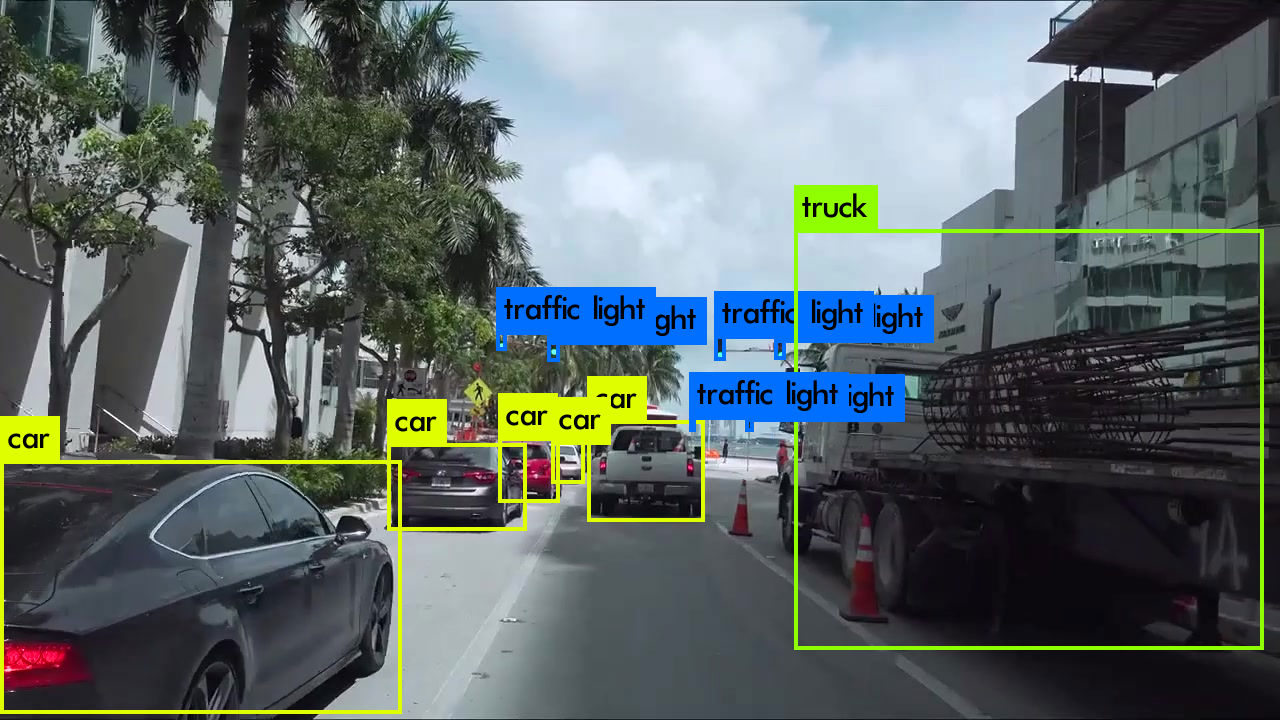}
\caption{Example of a misdetected truck with a large area {\color{black}using YOLO32full (left) and YOLO16full (right).}}
\label{fig:example_truck}
\end{figure*}

We observe that the total number of misdetections increases for vehicles with an area smaller than 0.006, which are considered to be small objects. 
In particular, misdetections for small objects are typically one order of magnitude lower than total detections, whereas misdetections for large objects are two orders of magnitude lower. Hence, we can conclude that, the larger the object, the lower the probability of experiencing a misdetection.

Regarding reported misdetections for large area objects, we have inspected visually a number of them, and in all cases, we have found scenarios where arguably no misdetection occurred in practice. For instance, Figure~\ref{fig:example_truck} shows an example of a truck with large area and a confidence level of 0.98 that is considered to be a misdetection. In this case, both configurations detect the truck, but the bounding box $IoU$ is below 0.5 because YOLO32full (left picture) includes the building behind the track in the bounding box. Hence, this YOLO16full prediction (right picture) is regarded as both FP and FN. In this particular example, YOLO16full predicts the object area more accurately than YOLO32full, but the opposite can also happen.

\begin{figure}[t!]
	\centering
	\includegraphics[width=1.0\columnwidth]{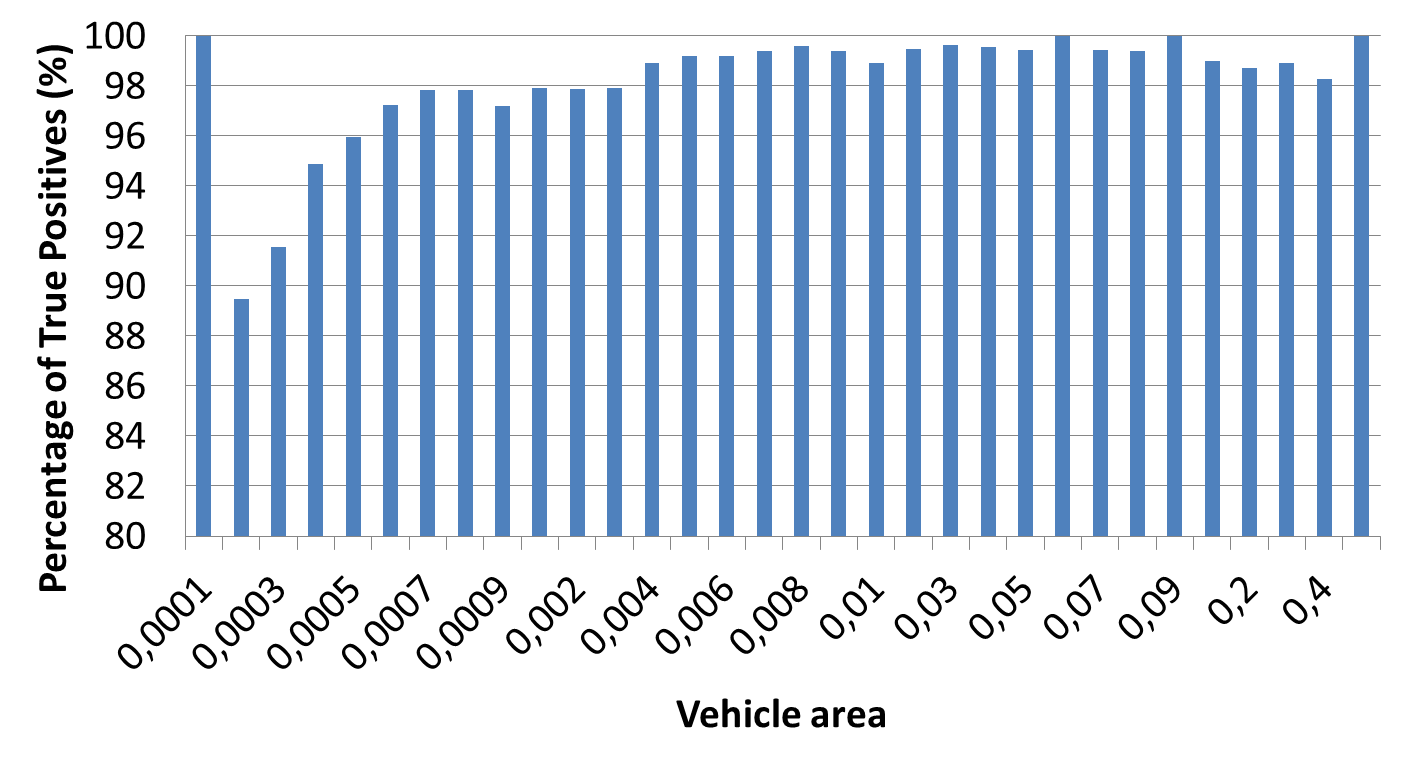}
	\caption{Percentage of correct detections (True Positives) for the YOLO16Approx $T_i$ video use case.}
	\label{fig:tp-yolo16app}
\end{figure}

The percentage of correct detections (TP), as shown in Figure~\ref{fig:tp-yolo16app}, decreases with object areas lower than 0.003, and decreases more steeply below 0.0006. Note that the smallest observed area has a 100\% of TP, but there are only 3 objects of this size, as shown in Figure~\ref{fig:area-misdet-yolo16app}. Therefore, this specific percentage can be simply disregarded since, with a larger number of objects within this area range, one would expect to observe a lower percentage of TP than for larger areas.

Overall, these results show that misdetections come mainly from objects with a small area, and {\color{black}have, therefore, low importance} in this context. Misdetections of objects of large area are mainly due to both configurations detecting the same object, but with the $IoU$ between the bounding-boxes being lower than the set threshold, and hence, the detections are regarded as FP and FN.

\subsubsection{Misclassification of Vehicle Types}
By visual inspection, we noticed that, in some cases, different configurations detect the same object but classify it differently. For example, one configuration may classify a vehicle as a car and another configuration may classify it as a truck, especially in the case of vans, as in this YOLO model there is no van object class. Misclassification of an object may have a significant impact in the mAP metric, since a misclassified object in these circumstances is regarded as both FP and FN, as the model failed to detect (classify in this case) an object, and detected a new object (in this case the same object but of a different object class).

To assess the impact of vehicle type misclassifications, we have merged the different vehicle classes present in the baseline YOLO model (car, motorbike, truck, and bus) into a single class called \emph{Vehicle}. {\color{black} We refer to this configuration as the \emph{Generalization} configuration.}

\begin{figure}[t!]
	\centering
	\includegraphics[width=1.0\columnwidth]{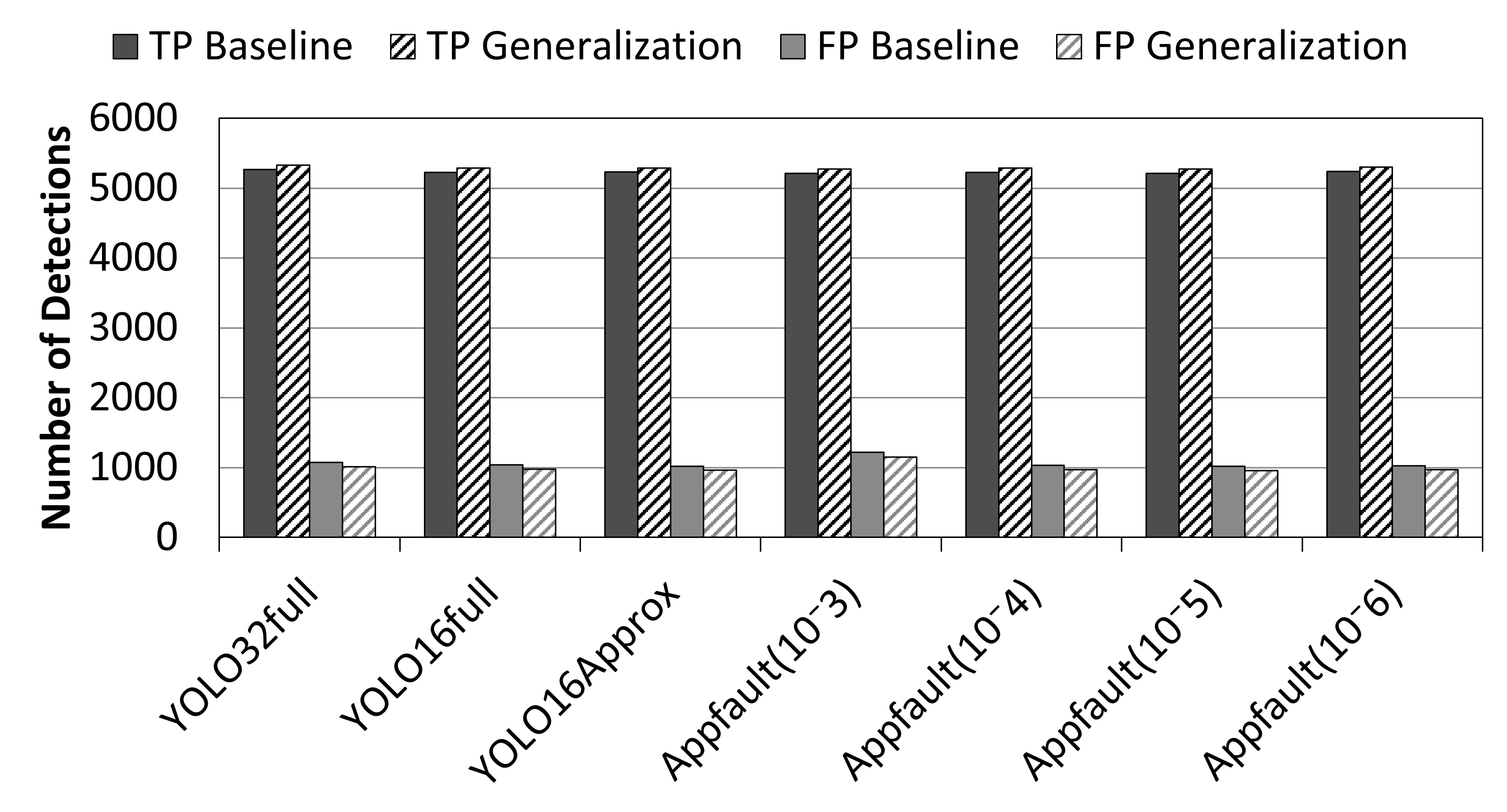}
	\caption{Comparison of the total number of T/P and F/P for the baseline and the generalization configurations for the labelled dataset.}
	\label{fig:tp-fp-gen-labelled}
\end{figure}

\begin{figure}[t!]
    \centering
    \begin{minipage}[t]{0.45\textwidth}
        \centering
        \includegraphics[width=1\textwidth]{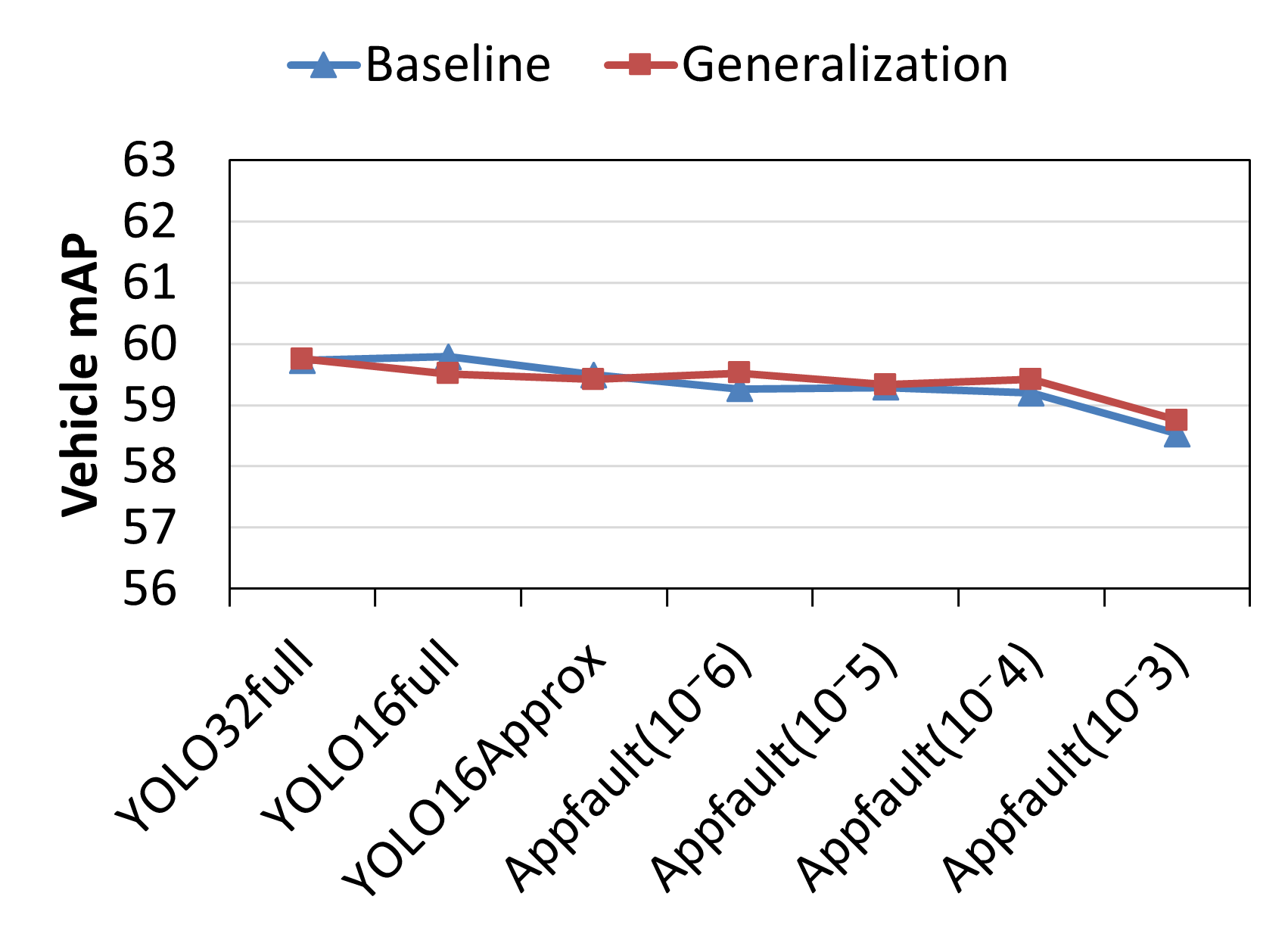}
        \caption{Comparison of the Vehicle mAP of the baseline and the generalization configurations for the labelled dataset.}
        \label{fig:map-gen-labelled}
    \end{minipage}\hfill
    \begin{minipage}[t]{0.45\textwidth}
        \centering
        \includegraphics[width=1\textwidth]{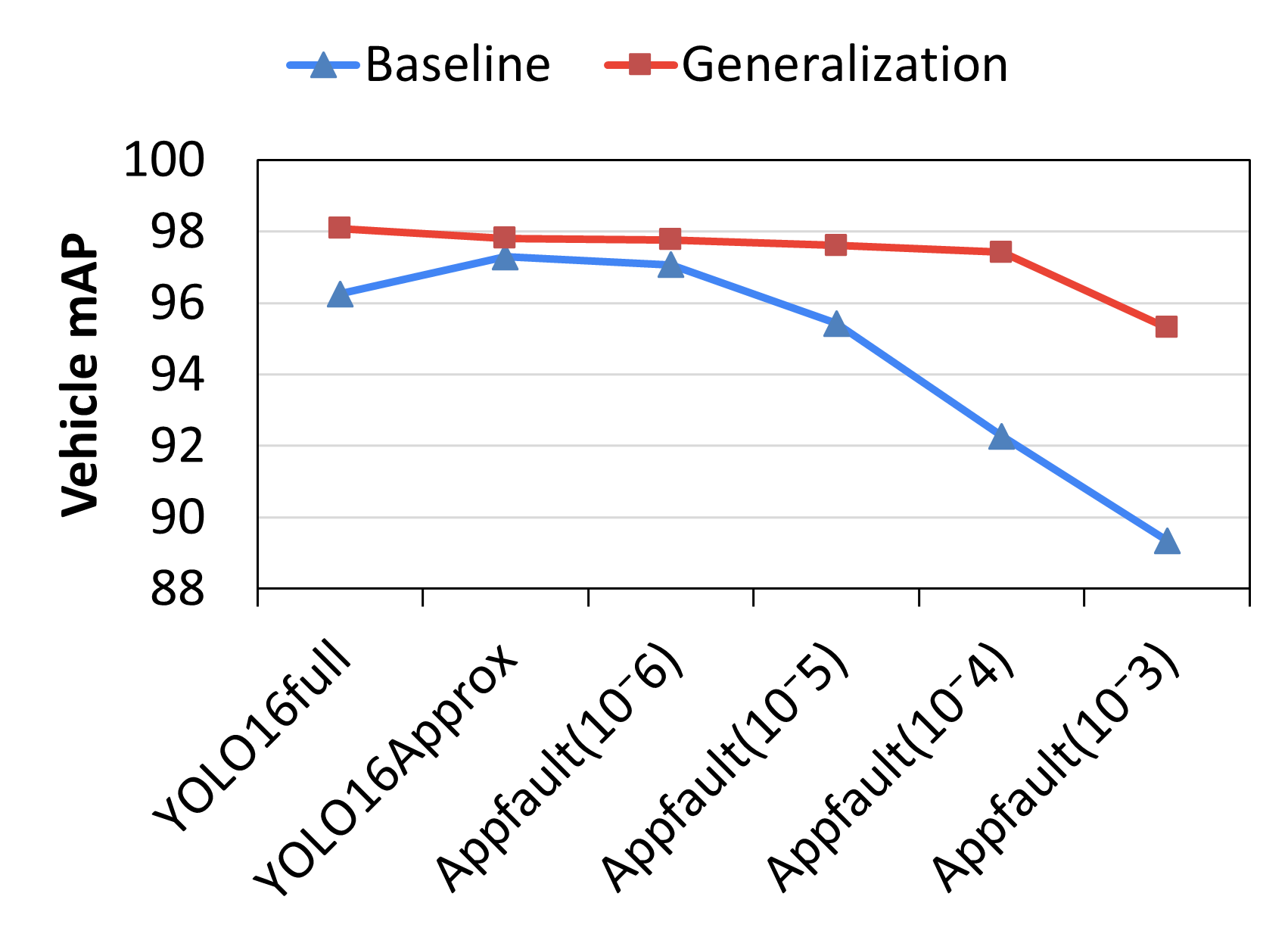}
        \caption{Comparison of the Vehicle mAP of the baseline and the generalization configurations for the $T_i$ video use case.}
        \label{fig:map-gen-ti}
    \end{minipage}
\end{figure}

Figure~\ref{fig:tp-fp-gen-labelled} shows the number of TP and FP of all vehicle types for the labelled dataset. In all configurations, the number of TP of the generalization increases a bit and the number of FP decreases by the same amount. Therefore, the accuracy of the object detector in these terms improves. On the other hand, the results for the mAP of \emph{Vehicles} in Figure~\ref{fig:map-gen-labelled}, show that the mAP of \emph{Vehicles} with the generalization decreases for some configurations (YOLO16full and YOLO16Approx), therefore contradicting to some extent the previous observation of increased accuracy. This shows a weak point of the mAP metric. In particular, the calculation of the mAP of \emph{Vehicles} for the baseline is the simple mean of the AP for the different types of vehicles without considering the number of observations of each class, whereas the AP of the generalization of vehicles is computed as a single value naturally weighting all objects. This implies that, if a type of vehicle has a very high/low AP but it accounts for a tiny fraction of the total objects with respect to the number of detections of other types of vehicles, the mAP of \emph{Vehicles} can be significantly affected.

For the video use case, we also obtain that the number of TP of the generalization increases and the number of FP decreases. On the other hand, in this case the mAP of the vehicle generalization provides a higher mAP for all configurations, as shown in Figure~\ref{fig:map-gen-ti} for the $T_i$ configuration. The $iT_i$ configuration follows analogous trends but it is omitted since it does not provide further insights.
{\color{black}A careful analysis of the videos shows that, if we do not generalize and keep different types of vehicles in different classes, YOLO32full performs a relevant number of misclassifications. Hence, it is often the case that approximate configurations perform a correct object class classification whereas YOLO32full does not. Assuming that approximate configurations are always wrong upon a mismatch with YOLO32full is, therefore, highly pessimistic.
However, since videos are not labelled, there is no automatic way to correct this issue. Generalizing vehicles into a single class is a way to mitigate this issue related to the lack of labelled driving videos.}

From the analysis of both labelled and unlabelled data, we can conclude that different configurations in some cases detect the same object but with a different classification, thus introducing more variability in the results if the mAP is solely analysed. 
{\color{black}
By considering all vehicle objects as a single class, we avoid issues related to unlabelled data misclassification by the baseline configuration. Moreover, in many cases -- yet not necessarily always -- classifying a vehicle object into the wrong subclass (e.g., classifying a car as a van) has irrelevant semantic impact.
}

\section{Conclusions}
\label{sec:concl}

Object detection in AD, is a stochastic process building upon deep learning, thus causing false positives/negatives. 
To tolerate errors, object detection is performed with sensor redundancy (multiple cameras, LiDARs, and radars), and time redundancy (leveraging detections over time), so that sporadic errors have no visible impact on the output.

This paper analyses the semantic impact of different approximation domains that can be used to save energy and complexity in the power-hungry CBOD process. In particular, {\color{black}in the context of an automotive case study,} we show how abundant but minor errors caused by lower precision and approximate arithmetic, as well as sporadic arbitrary errors caused by aggressive low voltage operation have, ultimately, negligible semantic impact in object detection despite causing some impact in the confidence level for detections and even some seldom misdetections. {\color{black}Our results for both, a labelled dataset and real driving videos, show that accuracy (in terms of mAP) is within 1\% of the original one even if we consider the three domains of approximation simultaneously decreasing floating-point precision from 32 to 16 bits, using approximate arithmetic, and considering fault rates of up to $10^{-5}$ in the mantissa of the data.}
{\color{black}Overall, our study provides strong ground to develop application-specific accelerators exploiting multiple approximation domains for cost-constrained domains such as the automotive one.}

\section*{Acknowledgments}
This work is partially funded by the DRAC project, which is co-financed by the European Union Regional Development Fund within the framework of the ERDF Operational Program of Catalonia 2014-2020 with a grant of 50\% of total cost eligible. 
This work has also been partially supported by the Spanish Ministry of Science and Innovation under grant PID2019-107255GB-C21/AEI/10.13039/501100011033. 

\bibliographystyle{unsrt}
\bibliography{biblio}

\end{document}